\documentclass[final,5p,times,twocolumn]{elsarticle}
\usepackage[english]{babel}
\usepackage{amsmath}
\usepackage{amssymb} 
\usepackage{graphicx}
\usepackage{booktabs}
\usepackage{multirow}
\usepackage{siunitx}
\usepackage{array}
\usepackage{pgffor}
\usepackage{here}
\usepackage{siunitx}
\usepackage{subcaption}

\journal{Computer Physics Communications}
\begin{document}
\title{Modifications in clusterization procedures for heavy-ion collisions: minimum spanning tree, simulated annealing, coalescence.}
\author{Viktar Kireyeu}
\affiliation{
Joint Institute for Nuclear Research, Joliot-Curie 6, 141980 Dubna, Moscow region, Russia
}

\date{\today}

\begin{abstract}
A new open-source cluster finding library ``Common Clusterization Library'' (CCL) is proposed to describe the clusters production when applied to the transport codes. The new library was applied to the Parton-Hadron-Quantum-Molecular Dynamics approach to be comparable with the established and well known MST and SACA implementations on the same set of the events. The presented procedures were tested with rapidity density distributions and complex observables like $\langle M_{imf} \rangle$ and $\langle Z_{\text{max}} \rangle$ vs $Z_{\text{bound}}$ at the energies of ALADIN and NA49 experiments. A modified simulated annealing chain were used to reduce the computational time. The new ``stable'' clusters tracking algorithm ``sMST'' for the ``lost'' QMD cluster recovering is presented. The results of the ``box-like'' coalescence procedure and a mixed method which incorporates the coalescence algorithm steps and the negative binding energy criteria are presented. The difference between the helium-3 found by the MST-based cluster finding procedure and the coalescence algorithm applied on the same set of events and baryons is shown for the first time.
\end{abstract}

\maketitle

\section{Introduction}
Decades ago transport approaches were proposed to address the problem of the heavy-ion collisions dynamics description (\cite{Aichelin:1985zz,Kruse:1985pg,Aichelin:1988me}). It is experimentally known that in such collisions initial nuclei can fragment into new single particles and heavier bound objects: nuclei and hypernuclei. One of the very first attempts to describe the cluster production was the application of the cluster finding procedure in \cite{Aichelin:1991xy}, which was based on the proximity in the coordinate space criterion -- so called ``Minimum Spanning Tree'', ``MST'' procedure \cite{zbMATH02560699}.

Then, a new method for the clusters recognition at earlier time steps, when particles (and bound groups of particles) are not yet well separated was proposed in \cite{Dorso:1992ch}, where authors firstly applied the Simulated Annealing (SA) procedure \cite{SA} to the transport approaches. The proposed method used the overall system binding energy as the minimization criteria for the Metropolis algorithm \cite{METROPOLIS}, but was presented for the relatively small systems.

The next evolution of the SA-based technique, named Simulated Annealing Clusterization Algorithm (SACA), was proposed in \cite{Puri:1998te} for heavier initial nuclei (larger systems): the authors used the classical SA procedure as the first ``Metropolis step'', then a second ``Fragment exchange'' step was applied. At this step fragments or nucleons obtained after the first step were passed again through the SA procedure to overcome the trapping to any local minima.

Recently, a new approach for the early clusters identification in the transport approaches, FRIGA, was described in \cite{LeFevre:2019wuj}. It was demonstrated that inclusion of additional components into the binding energy calculations yields into the better description of the isotopes and hypernuclei production.

All proposed approaches show good results and agrees with the experimental data in wide energy range, however all of them are closed-source and are available mostly for a very limited set of the transport codes if not heavily tied to the single one.

In order to overcome this limitation a new open-source library ``Common Clusterization Library'' (CCL) \cite{ccl_github} is developed on the base of previously presented ``psMST'' \cite{Kireyeu:2021igi} library. The CCL library is distributed under the terms of the ``Let's Be Scientific'' (``LBS'') license which allows for the free code usage and modification, but strictly requires the explicit permission by the authors prior any kind of results publication to avoid the inappropriate use of the code. The full license text is distributed with the library code.

The following procedures are implemented within CCL:
\begin{itemize}
    \item ``Box-like'' coalescence for light nuclei and hypernuclei.
    \item Cluster finding procedure based on the Minimum Spanning Tree with the usage of Disjoint-set data structure (DSU MST).
    \item Recovering of the lost QMD clusters during the reaction evolution.
    \item A new DSU MST and two-pass Simulated Annealing chain.
\end{itemize}

The MST-based clusterization will be described in Section \ref{sec:MST}. The Simulated Annealing procedures and their results are covered in Section \ref{sec:SA}. The coalescence approach and ``mixed'' procedure are presented in Section \ref{sec:Coal}.

The results of the cluster finding procedures (MST, SACA and CCL SA) will be tested with ALADIN experiment complex observables:
\begin{itemize}
    \item $\langle M_{\text{imf}} \rangle$ -- stands for the mean multiplicity of the forward-emitted ($\theta < \ang{90} $ or $y > 0$) intermediate mass fragments (IMF) with the charge $3 \leq Z \leq 30$.
    \item $\langle Z_{\text{max}} \rangle$ -- average maximal charge of the single fragment in the event.
    \item $Z_{\text{bound}}$ -- total number of bound charges ($Z \ge 2$) within the forward-emitted clusters in event.
\end{itemize}
Additional observable to control the clusterization results quality:
\begin{itemize}
    \item $\langle E_{\text{bind}} \rangle$ -- the mean binding energy of all forward-emitted clusters of the event.
\end{itemize}

The coalescence and the ``mixed'' procedure, which combines the coalescence algorithm steps and the usage of the $E_{\text{bind}}$ criterion, will be shown with the NA49 experimental data to be comparable with previous coalescence results within ``psMST'' \cite{Kireyeu:2023spj}. Only QMD-based approaches will be used as base models to simulate the data as in contrast to BUU-based models they can propagate particles correlations through the collision evolution without additional modifications \cite{LeFevre:2019wuj}. In present work, the Parton-Hadron-Quantum-Molecular Dynamics (PHQMD) \cite{Aichelin:2019tnk} transport approach was used to generate the data sets.

\section{MST, DSU}
\label{sec:MST}
The spanning tree is a simplified version of the connected graph: it is a subgraph that uses the original graph's vertices but only enough edges to connect every vertex without cycles. A single graph can have many different spanning trees.
When each edge in the original graph has a ``cost'' or ``weight'', we can look for the spanning tree with the smallest total ``weight'' -- this is called a Minimum Spanning Tree (MST).

For the given set of $N$ particles in the event $\{ p_i \}_{i=1}^N$, the graph $G = (V, E)$ is defined with:
\begin{itemize}
    \item Vertices: $V = \{ p_1, p_2, \dots, p_N \}$,
    \item Edges: $E = (p_i, p_j)$.
\end{itemize}

The MST-based procedure is adapted for particle clustering, where particles are treated as vertices, and edges represent proximity relationships. The weights of the edges implicitly defined by a distance threshold (``clusterization radius'' $r_{clust}$) in the $(i,j)$ pair center-of-mass (COM) frame:
\begin{equation}
r_{\text{clust}} = \sqrt{(x_i' - x_j')^2 + (y_i' - y_j')^2 + (z_i' - z_j')^2}
\end{equation}
Here $(x_i', y_i', z_i')$ and $(x_j', y_j', z_j')$ denote the spatial coordinates of particles $i$ and $j$ after Lorentz transformation to their two-particle center-of-mass frame. Usually, the ``MST'' in the context of the ``clustering'' in the heavy-ion collisions is understood as an algorithm which relies on the coordinate space only, however, in the authors opinion, it is not limited to this and could therefore be extended by including the momentum space too or any other physically justified criteria. Thus, a possibility to check the proximity in momentum space for baryons in the center‑of‑mass frame of the checked pair or cluster is also available. For a pair, the two baryons are boosted to their pair COM frame and the pair is accepted only if each baryon satisfies condition $|\mathbf{p}_i'| < \Delta p$. For a cluster, the merged system is boosted to its COM frame and the merge is accepted only if all constituents satisfy $|\mathbf{p}_i'| < \Delta p$.
Here $\mathbf{p}_i'$ is the momentum of baryon $i$ after Lorentz transformation to the relevant center-of-mass frame (pair COM for a two-particle check, or cluster COM for a cluster check). The parameter $\Delta p$ here is the upper cut on the absolute momentum of each constituent in this frame.

The Disjoint Set Union (DSU), also known as Union-Find, is a data structure that efficiently manages a partition of a set of elements into disjoint (non-overlapping) subsets, for the first time described in \cite{DSU}. It supports two primary operations:
\begin{itemize}
    \item Find: determines the representative (or ``root'') of the subset containing a given element. Two elements are in the same set if and only if their ``Find'' operations return the same representative.
    \item Union: merges two subsets into a single subset.
\end{itemize}

For the given event with $N$ particles and a ``roots'' array $S$ the DSU structure manages:
\begin{enumerate}
    \item Initialization: $S[i] = i, \quad \forall i \in \{0, 1, \dots, N-1\}.$

    \item Find Operation:
    \begin{equation}
        \text{find}(i) = \begin{cases} 
        i & \text{if } i = S[i], \\
        \text{find}(S[i]) & \text{otherwise},
        \end{cases}
    \end{equation}
    with path compression: $S[i] \leftarrow \text{find}(S[i])$.

    \item Union Operation: for particles $i, j$:
    \begin{itemize}
        \item Compute roots: $r_i = \text{find}(i)$, $r_j = \text{find}(j)$,
        \item If $r_i \neq r_j$: $S[r_j] = r_i$.
    \end{itemize}
\end{enumerate}
DSU is particularly efficient for problems involving grouping, such as constructing the MST clusters in a graph. The path compression, which is also included in CCL, enhance the ``find'' operation, thus the querying the ``root'' (cluster) of a particle is nearly constant time after initial queries.

\begin{figure}[ht]
    \includegraphics[width=\linewidth]{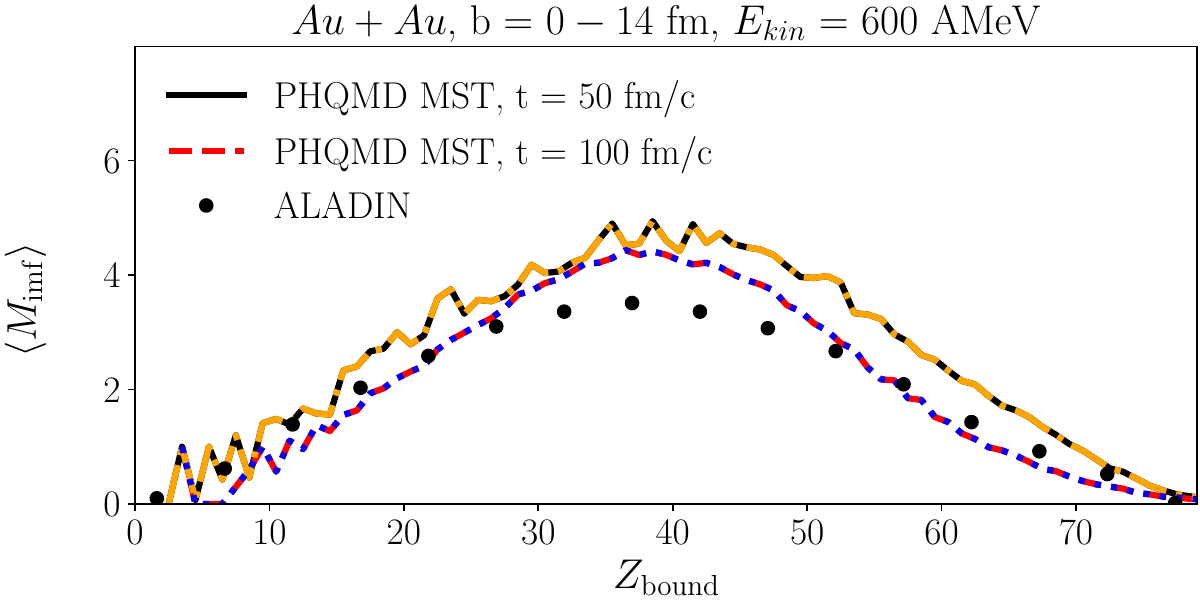}
    \includegraphics[width=\linewidth]{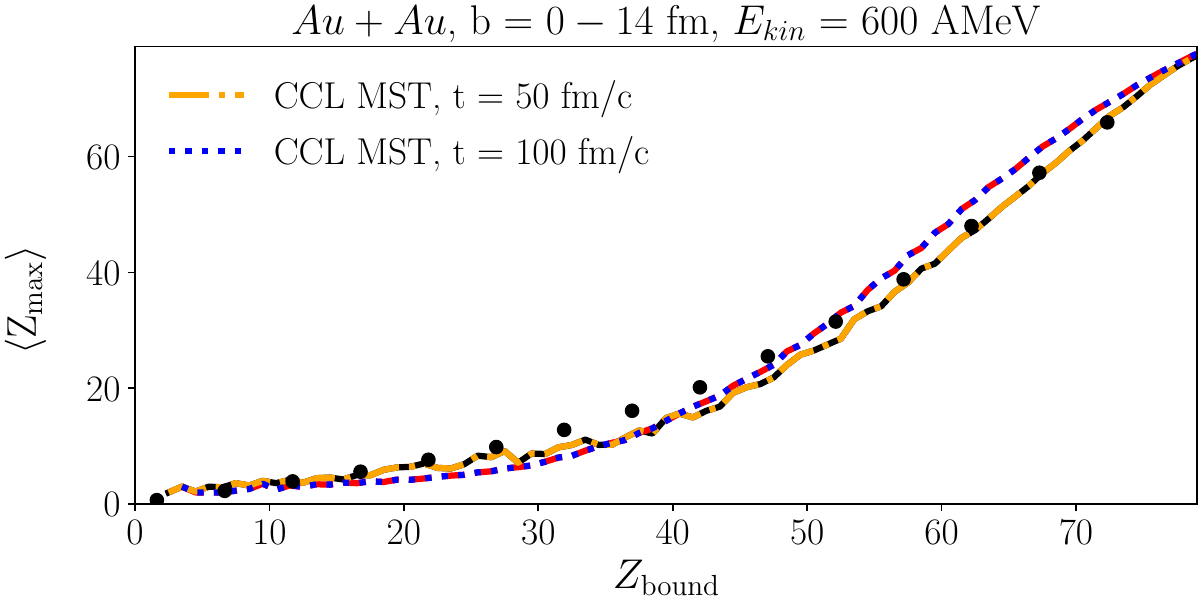}
    \caption{Upper panel: $\langle M_{imf} \rangle$ vs $Z_{\text{bound}}$; Lower panel: $\langle Z_{\text{max}} \rangle$ vs $Z_{\text{bound}}$; Black solid lines -- PHQMD MST at $t = 50$ fm/c. Red dashed lines -- PHQMD MST at $t = 100$ fm/c. Orange dash-dotted lines -- CCL MST at $t = 50$ fm/c. Blue dotted lines -- CCL MST at $t = 100$ fm/c. ALADIN experimental data (black circles) are taken from \cite{Schuttauf:1996ci,LeFevre:2019wuj}.}
    \label{fig:mst}
\end{figure}

The MST procedure in CCL is implemented as DSU based, while MST in PHQMD constructs the full graph. Fig. \ref{fig:mst} show the comparison of the MST clusterization results in min. bias $Au + Au$ collisions at $E_{kin} = 600$ AMeV: $\langle M_{imf} \rangle$ (upper panel) and $\langle Z_{max} \rangle$ (bottom panel) vs $Z_{\text{bound}}$ for both methods applied on the same set of events at two time steps: $t = 50$ fm/c and $t = 100$ fm/c. All observables: $\langle M_{imf} \rangle$, $\langle Z_{max} \rangle$ are calculated for the forward-produced clusters.

It is clearly visible, that both MST implementations provides identical results.

This agreement is expected for the pure coordinate-space MST case.
For a fixed event and a fixed clusterization radius, the MST clustering is deterministic: all particle pairs satisfying the same proximity criterion define the same connected components of the graph. Therefore, if the built-in PHQMD MST and the CCL MST are applied to exactly the same particle sample, they must return identical clusters up to the ordering of particles and clusters. The technical difference is only in the graph construction. The PHQMD implementation builds the full graph explicitly, whereas the CCL implementation uses the DSU (Union-Find) structure to merge connected components directly. This reduces the computational time while preserving the same connectivity criterion and therefore the same physical result.

As it is seen on the figure, the MST-based clustering overestimate the mean multiplicity of the intermediate mass fragments and finds mostly unbound clusters at earlier time steps.

\begin{figure}[ht]
    \includegraphics[width=\linewidth]{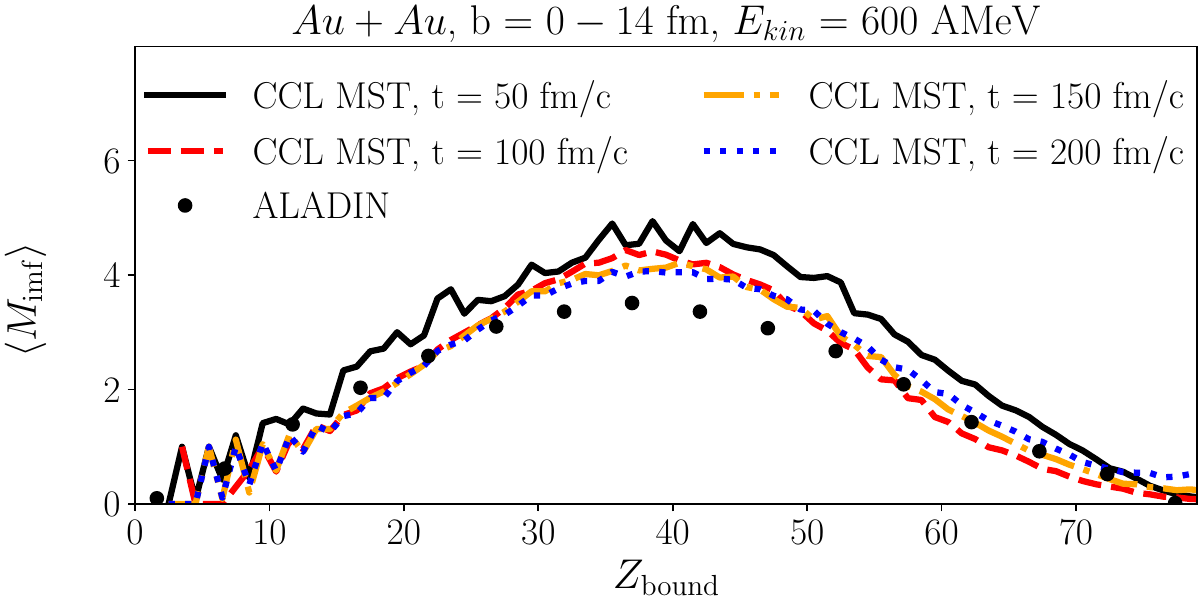}
    \includegraphics[width=\linewidth]{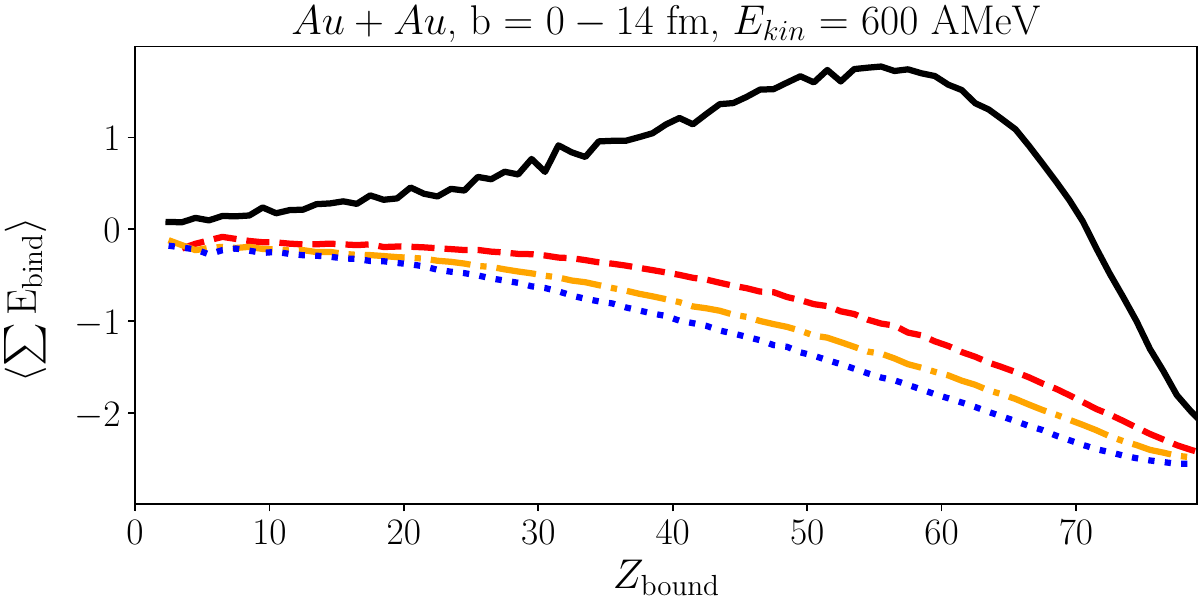}
    \caption{$\langle M_{imf} \rangle$ (top panel) and $\langle E_{\text{bind}} \rangle$ (bottom) vs $Z_{\text{bound}}$ for CCL MST. Black solid lines -- $t = 50$ fm/c, red dashed lines -- at $t = 100$ fm/c, orange dash-dotted lines -- $t = 150$ fm/c, blue dotted lines -- $t = 200$ fm/c. Black circles -- ALADIN experimental data \cite{Schuttauf:1996ci,LeFevre:2019wuj}.}
    \label{fig:ccl}
\end{figure}

The upper panel of Fig. \ref{fig:ccl} show the CCL MST clusterization results for four time steps: $t = 50,\, 100,\, 150,\, 200$ fm/c. The calculated mean binding energy $\langle E_{bind} \rangle$ is also plotted on the bottom panel. While the ``rise and fall'' curve is stably reproduced at all time steps, the mean binding energy at the earlier time is mostly positive for lighter IMF clusters, that means they are not actually ``bound''. Same picture is also true for the original MST procedure.

\section{SACA, SA}
\label{sec:SA}
To overcome the problem of the overestimation of the experimental points, another procedure was proposed to be applied on top of the MST clusters -- the Simulated Annealing Clusterization Algorithm (SACA). In short, the main idea of the SACA procedure was to take the MST pre-fragments and free baryons as an input and then run the two stages of the minimization procedure for the total binding energy in the system by recombination of bound and unbound baryons in all possible ways.

The main problem with the total system binding energy check is related to the combinatorial issue. For the $Au + Au$ system with 394 baryons the number of all possible combinations of free and bound particles is represented by the Bell number $B_{394}$ which is roughly $B_{394} > 10^{500}$ combinations per event. Despite the fact that the Simulated Annealing procedure is limited by the control parameters such as ``temperature'', ``cooling rate'', ``number of tries'', and the ``stagnation'' counter to speedup the calculational time, and thus can not check all available combinations, it can find the system binding energy close to the global minimum.

The CCL library also provides the classical implementation of the Simulated annealing procedure (``CCL SA''). The CCL SA algorithm operates on a set of particles, each particle can be assigned to the cluster or be free (unbound), and iteratively proposes new cluster configurations by:
\begin{itemize}
    \item moving particle from cluster to the new singleton (particle became ``free''),
    \item transferring free particle, or particle bound in one cluster to another existing cluster.
\end{itemize}
The new system is always accepted if its total binding energy is lower than binding energy of the previous system: $\Delta E_{bind} = E_{bind_{proposed}} - E_{bind_{current}} < 0$. The escape from the local minimum can be governed by the Metropolis-Hastings criterion, which allows to occasionally accept higher-energy configurations: $P_{\text{accept}} = \exp\!\big(-\Delta E / T\big)$. In the next subsection the algorithm will be described in more details.

\subsection{CCL SA description}
The CCL Simulated Annealing (CCL SA) procedure consists of two stages: let's call them ``pass 1'' and ``pass 2''.

The ``pass 1'' procedure is the classical implementation of the Simulation Annealing algorithm which optimizes particle clustering by minimizing the total binding energy $E_{\text{bind}}$ of a system of $N$ particles $\{ p_i \}_{i=1}^N$, each with a four-momentum $(m_i, \mathbf{p}_i)$ and position $(t_i, \mathbf{x}_i)$, and an initial cluster assignment from a prior MST clustering. The ``all baryons are free'' or ``one single cluster'' inputs are also allowed, the latter is subsequently used within the combined SA chain in CCL.

During the ``pass 2'' stage, clusters and free particles after the ``pass 1'' are recombined in that way, that baryons can not be taken out of the clusters: baryons and whole clusters can only merge into bigger clusters. The $E_{\text{bind}}$ minimization criterion is the same as during the ``pass 1''.

\subsubsection{System Binding Energy}
The total binding energy of the system is defined as:
\begin{equation}
E_{\text{system}}(\mathcal{S}) = \sum_{C \in \mathcal{S}} |C| \cdot E_{\text{bind}}(C), \quad \text{for } |C| > 1,
\end{equation}
where $\mathcal{S} = \{ C_1, C_2, \dots, C_k \}$ is the set of clusters, $|C|$ is the number of particles in cluster $C$, and $E_{\text{bind}}(C)$ is the binding energy \textit{per nucleon} of cluster $C$, computed based on the four-momenta of its constituent particles in the cluster rest frame.

\subsubsection{Algorithm Control Parameters}
The algorithm uses the following parameters:
\begin{itemize}
    \item Initial temperature: $T = T_{\text{max}}$.
    \item Minimum temperature: $T_{\text{min}}$.
    \item The cooling rate $\alpha \in (0, 1)$ controls the temperature update after each temperature level:
        \begin{equation}
        T \leftarrow \alpha T
        \end{equation}
        In the present calculations $\alpha = 0.95$ was used.
    \item Steps mode: determines the number of iterations per temperature level:
    \begin{equation}
    \text{steps} = \begin{cases} 
    S & \text{if fixed mode}, \\
    S \cdot \frac{T}{T_{\text{max}}} & \text{if linear mode}, \\
    S \cdot \alpha^{\ell} & \text{if exponential mode},
    \end{cases}
    \end{equation}
    where $S$ is the base number of steps, and $\ell$ is the temperature level index.
    \item Probability of creating a new singleton: $P_{\text{new}}$, optionally adjusted linearly:
    \begin{equation}
    P_{\text{new}} = P_{\text{new,min}} + (P_{\text{new,max}} - P_{\text{new,min}}) \cdot \frac{T - T_{\text{min}}}{T_{\text{max}} - T_{\text{min}}}
    \end{equation}
    where $P_{\text{new,max}}$ and $P_{\text{new,min}}$ define the probability of proposing a move-to-singleton trial at the initial and final temperature limits, respectively. This adjustment leads to the higher probability for a particle to became free at earlier ``temperature'', and to be bound into the existing cluster at the lower levels when the ``temperature'' cools down.
    \item Switch to enable or disable the usage of the Metropolis-Hastings criterion during minimization procedure.
\end{itemize}

\subsubsection{Algorithm Workflow}
During the SA procedure, a configuration $\mathcal{C}$ denotes a complete partition of all input baryons into clusters and singletons. The current configuration is denoted as $\mathcal{C}_{\text{curr}}$, and the proposed configuration after a trial move is denoted as $\mathcal{C}_{\text{prop}}$. An individual cluster inside a configuration is denoted by $C$. Within a trial move, $C_{\text{src}}$ denotes the source cluster from which a baryon is selected, and $C_{\text{dst}}$ denotes the destination cluster to which this baryon is moved.

During ``pass 1'', at every trial a candidate state $\mathcal{C}_{\text{prop}}$ is drawn from $\mathcal{C}_{\text{curr}}$ with one of two elementary moves:

\paragraph{(A) Move to singleton (free particle):}
Select uniformly a non-singleton cluster $C_{\text{in}}$ and uniformly select a particle $p \in C_{\text{in}}$, move it out of the cluster and set $C_{\text{in}} \leftarrow C_{\text{in}} - \{p\}$. 

\paragraph{(B) Move between clusters (including singletons):}
Choose two different clusters $C_{\text{src}}$ and $C_{\text{dst}}$ uniformly at random (they may be singletons). Choose a particle $p \in C_{\text{src}}$ uniformly (if $|C_{\text{src}}|=1$ then $p$ is that singleton).  
Perform
\begin{equation*}
C_{\text{src}} \leftarrow C_{\text{src}} - \{p\},\qquad
C_{\text{dst}} \leftarrow C_{\text{dst}} + \{p\}.
\end{equation*}
If $C_{\text{src}}$ becomes empty, remove it. \par

At the the ``pass 2'' stage only the move $(B)$ is allowed.

\begin{table*}[!t]
\centering
\begin{tabular}{lcccccccccc}
\toprule
Label & $r_{\text{clust}}$ & $\Delta p$ & $P_{new,max}$  & $P_{new,min}$ & $T_{\text{max}}$ & $T_{\text{min}}$ & $\text{Steps}_{\text{ pass }1}$ & $\text{Steps}_{\text{ pass }2}$ & $s_\text{min}$ & $s_\text{den}$\\
\midrule
S1 & 2.5 & 0.3 & 0.9 & 0.1 & 1 & $10^{-15}$ & 500 & 4000 & 50 & 2\\
S2 & 2.5 & --  & 0.9 & 0.1 & 1 & $10^{-15}$ & 500 & 4000 & 50 & 2\\
\bottomrule
\end{tabular}
\caption{CCL Simulated Annealing procedure parameters for ``pass 1''. $r_{\text{clust}}$ is in 'fm' and $\Delta p$ are in 'GeV/c' units.} 
\label{tab:sa_param}
\end{table*}

\subsubsection{Acceptance criterion}
Let the total binding energies of $\mathcal{C}_{\text{curr}}$ and $\mathcal{C}_{\text{prop}}$ be
\begin{equation}
E_{\text{curr}} = \sum_{C \in \mathcal{C}_{\text{curr}}} |C|\,E_{\text{bind}}(C),
\qquad
E_{\text{prop}} = \sum_{C \in \mathcal{C}_{\text{prop}}} |C|\,E_{\text{bind}}(C).
\end{equation}

Define
\begin{equation}
\Delta E = E_{\text{prop}} - E_{\text{curr}}
\end{equation}

The acceptance rule is:
\begin{itemize}
    \item $\Delta E < 0$: always accept.
    \item $\Delta E \ge 0$: accept if $u < \exp(-\Delta E / T)$, where $u$ is a random number sampled from the uniform distribution on the interval $[0,1)$.
\end{itemize}

When Metropolis--Hastings (MH) acceptance criterion is enabled the algorithm accepts ``uphill'' moves with probability $\exp(-\Delta E/T)$, allowing escape from local minimum.

The early exit from the SA loop is handled by the ``stagnation'' counter.
Let 
\begin{equation}
L = \max\!\left(s_\text{min}, \frac{\text{steps}}{\max(1, s_\text{den})} \right)
\end{equation}
where $s_\text{min}$ is the minimum stagnation length, $s_\text{den}$ controls the temperature-level dependent limit through $\text{steps}/s_\text{den}$, and $c$ is the counter of consecutive rejected moves. Then, for each rejected move: $c \leftarrow c + 1$; for each accepted move: $c \leftarrow 0$. The simulated annealing loop is terminated early if $c \ge L$.

\subsection{Full CCL SA chain}
The full ``CCL SA'' clusterization chain consists of:
\begin{enumerate}
    \item MST stage: at this stage pre-fragments are found by the proximity in the coordinate space (and possibly in the momentum space too) criteria. At the earlier time steps these found pre-fragments can be not bound enough. At this stage three arrays are filled:
    \begin{itemize}
        \item Free (unbound) baryons.
        \item ``stable'' clusters with $E_{\text{bind}}/A < E_{\text{cut}}$.
        \item ``Unstable'' clusters  with $E_{\text{bind}}/A > E_{\text{cut}}$.
    \end{itemize}
    Here $E_{\text{cut}} = -4$ MeV. This value is used as a simple technical separation criterion rather than as a new physical constant. It was chosen as an intermediate value between the typical binding energies per nucleon of weakly bound light nuclei and the more strongly bound medium/heavy nuclei.

    \item SA ``pass 1'': at this step each of the previously found MST-based ``unstable'' clusters with $E_{\text{bind}}/A > E_{\text{cut}}$ MeV is passed through the Simulated Annealing procedure to find most bound configuration within this cluster. As it is only single cluster which is passed to the SA each time,  it is more profitable to increase the probability of a baryon to became ``free'' on this stage to find more smaller clusters and then preform ``pass 2''. At the end of the ``pass 1'' clusters with the binding energy $E_{\text{bind}}/A > E_{\text{cut}}$ are deconstructed to the unbound baryons, then newly found ``free'' baryons (if any) are passed to the unbound baryons array, clusters with $E_{\text{bind}}/A < E_{\text{cut}}$ are stored in the ``stable'' clusters array. This approach is new for the SA-based cluster finding procedures. It reduces the computational time by applying the first SA pass to individual unstable MST pre-fragments instead of to the full event. As a result, the number of trial moves per temperature level can be significantly reduced in pass 1.

    \item SA ``pass 2'': during this step clusters are not anymore allowed to ``decay'', a baryon can not be taken out of a cluster. All clusters, ``stable'' $E_{\text{bind}}/A < E_{\text{cut}}$, ``unstable'', and free baryons are recombined to get the most bound configuration: baryon can be added to another baryon to form a new cluster, baryon can be added to the existing cluster or one existing clusters can be added to another cluster to form a bigger fragment. 
\end{enumerate}

Steps 1 and 2 are the key difference between this procedure and the original SACA \cite{Puri:1998te} idea, where the MST clusters plays the role of 'seeds' for the first pass of the Simulated Annealing applied on the whole set of the MST clusters and free baryons.

It is more profitable in terms of the calculational time to use less initial steps for the ``pass 1'' and increase it for the ``pass 2''. The ``stagnation''-governed early exit reduce the calculational time even more.

For this study the ``CCL SA'' procedure was applied to the simulated data with the parameters shown in the Table \ref{tab:sa_param}.

\subsection{SACA and SA results}
\begin{figure}[ht!]
    \includegraphics[width=\linewidth]{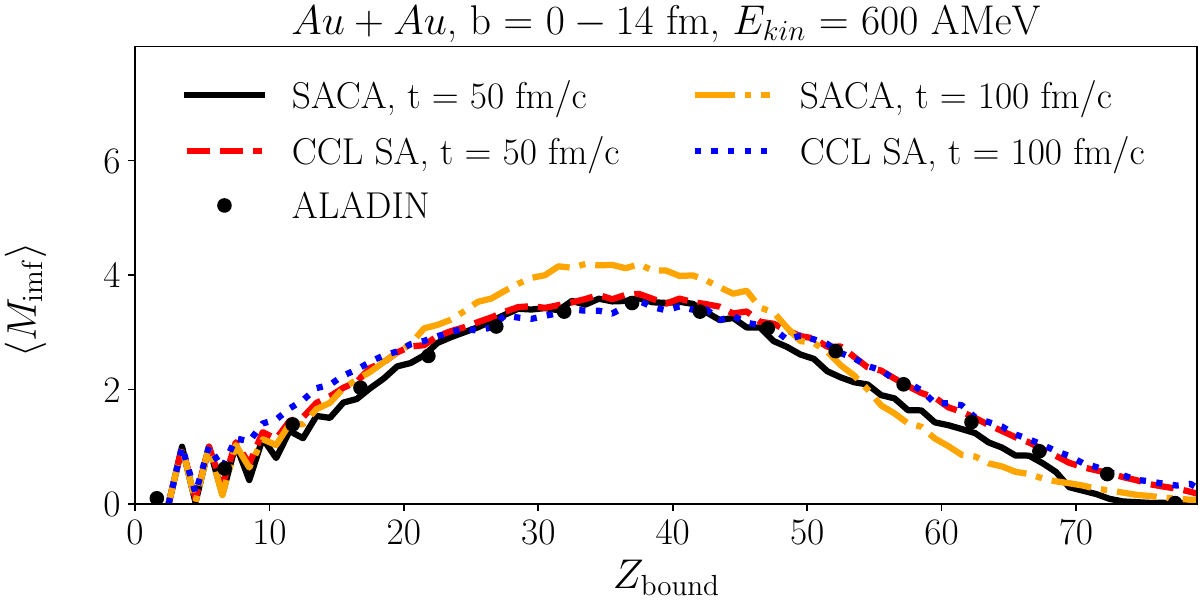}
    \includegraphics[width=\linewidth]{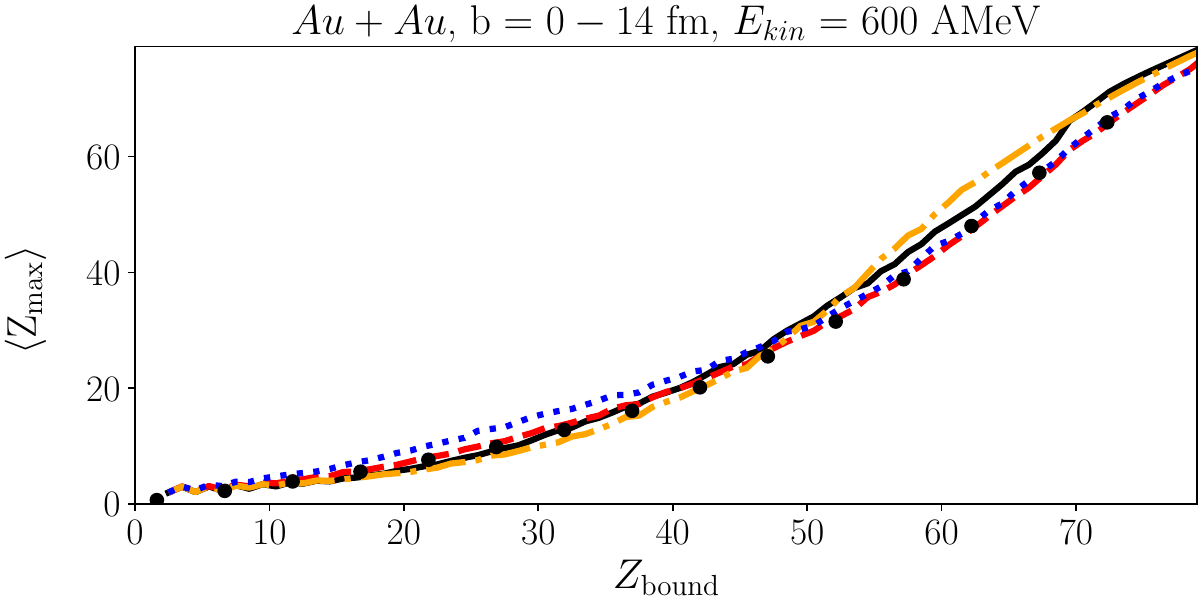}
    \includegraphics[width=\linewidth]{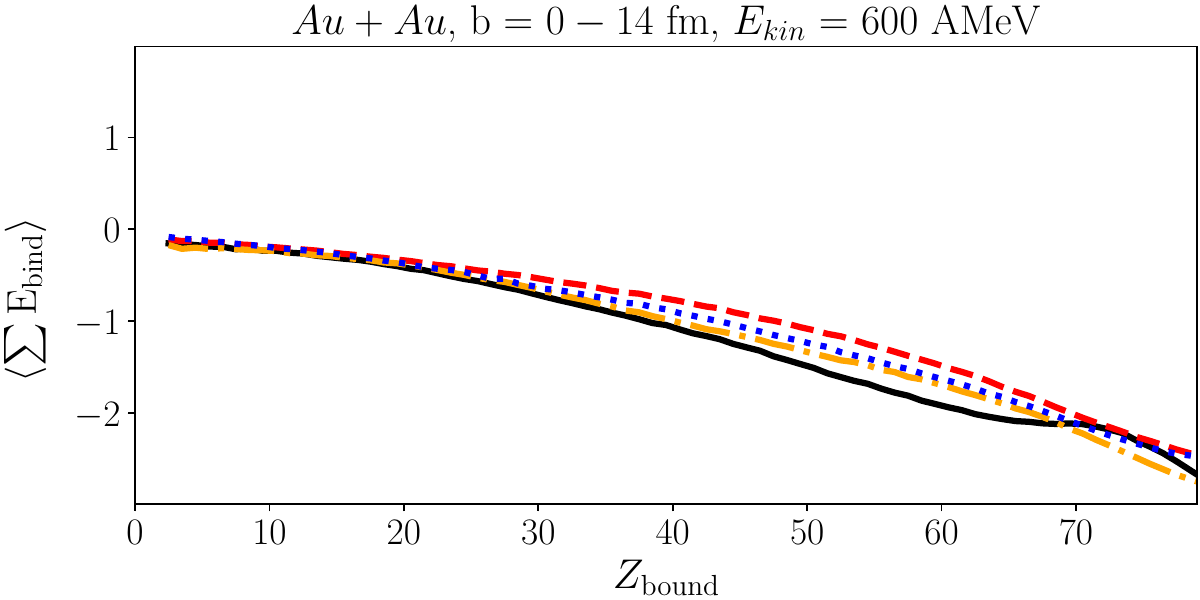}
    \caption{$\langle M_{imf} \rangle$ (top panel), $\langle Z_{\text{max}} \rangle$ (middle panel) and $\langle E_{bind} \rangle$ (bottom panel) as a function of $Z_{\text{bound}}$. Black solid lines and orange dash-dotted lines are PHQMD SACA at $t = 50$ and $t = 100$ fm/c. Red dashed lines and  blue dotted lines -- CCL SA at $t = 50$ and $t = 100$ fm/c. Black circles -- ALADIN experimental data taken from  \cite{Schuttauf:1996ci,LeFevre:2019wuj}.}
\label{fig:saca_vs_sa}
\end{figure}

The results of the PHQMD SACA and CCL SA procedures are shown the Fig. \ref{fig:saca_vs_sa} for the same events at the time $t = 50$ and $t = 100$ fm/c. At earlier time $t = 50$ fm/c both procedures succeeds to describe the ALADIN experimental data, however, the proposed ``CCL SA'' routine with the ``S1'' preset shows a bit better agreement. The PHQMD SACA very slightly underestimates experimental $\langle M_{\text{imf}} \rangle$ points at larger $Z_{\text{bound}} > 45$. 

Similar behavior is true for the ``$\langle Z_{\text{max}} \rangle$ vs $Z_{\text{bound}}$'' observable: both SACA and SA curves are very close to the data points with slight overestimation from SACA at larger $Z_{\text{bound}} > 45$, but the ``CCL SA'' agreement with the data is perfect.

The mean binding energy of the system $\langle E_{\text{bind}} \rangle$ curves for both methods surprisingly differs: the curve of the ``CCL SA'' procedure is more monotonic, while SACA give some ``bump''-like structure at the medium to high $Z_{\text{bound}}$ region.

At the larger time $t = 100$ fm/c the ``CCL SA'' used the ``S2'' preset without proximity checks in the momentum space: at larger times baryons with the high relative momentum are mostly not at the same place. At this time step both methods again shows good agreement with the data, but SACA have a bit larger discrepancies: it overestimates $\langle M_{\text{imf}} \rangle$ at smaller $ 20 < Z_{\text{bound}} < 45$ and underestimates same quantity at  $Z_{\text{bound}} > 50$; it also overestimates $\langle Z_{\text{max}} \rangle$ at larger $Z_{\text{bound}} > 50$.

While ``CCL SA'' has some overestimation of the $\langle M_{\text{imf}} \rangle$ at smaller $ 20 < Z_{\text{bound}} < 45$ (however, smaller than SACA), it perfectly describes the tail of the distribution and the $\langle Z_{\text{max}} \rangle$ vs $Z_{\text{bound}}$ dependence.

As for the mean binding energy $\langle E_{\text{bind}} \rangle$, at later time step both methods have similar smooth curves which are close to each other.

``CCL SA'' results gives an insight: it is not enough to have some single fixed set of parameters in the MST of SA cluster finding approach to describe complex observables such as ``rise and fall'' of IMF -- e.g. at earlier stages it is more logical to use smaller $r_{\text{clust}}$ than at later times and/or to also check the proximity in the momentum space.

\section{Tracking}
\label{sec:track}
As it was discussed in \cite{Coci:2023daq}, the QMD dynamics propagates single baryons and if the MST-based clusterization procedure marks group of baryons as a stable cluster, on the next time step this cluster can dissolve without collision occurred even if the corresponding quantum system must be stable. The authors proposed the ``advanced MST'' (``aMST'') procedure, which uses the collision history and the information of the kinetic ``freeze-out'' time to reconstruct back some of the artificially ``destroyed'' clusters.

The CCL library has a different tracking technique which works sequentially from the first time slice till the last without the explicit knowledge of the kinetic ``freeze-out'' time, using only information about particles collisions within current and previous time steps.

Consider a set of input clusters
\begin{equation*}
\mathcal{P} = \{ p_1, p_2, \ldots, p_N \},
\end{equation*}
with each cluster \(p\) having a set of constituent particles:
$$
C(p) = \{ c_1, c_2, \ldots, c_{k(p)} \}
$$
A cluster $p$ from the input vector enters the stable candidates pool for further tracking only if it satisfies the physical stability criteria:
\begin{equation}
E_{\text{bind}}(p) < E^{\text{cut}}_{\text{bind}}
\qquad\text{and}\qquad
\forall\, c \in C(p):\;
t_{\text{coll}}(c) \leq t_{\text{step}},
\end{equation}
where $E^{\text{cut}}_{\text{bind}} \leq 0$ is a configurable threshold (default: any negative value),  $t_{\rm coll}(c)$ is the current latest collision time recorded for constituent baryon $c$, and $t_{\rm step}$ is the current evaluation time. Let $\mathcal{S} = \{ s_1, s_2, \dots, s_M \}$ denote the persistent list of currently registered stable clusters carried over from previous time steps with initial $\mathcal{S} = \varnothing$. Prior to processing new candidates, $\mathcal{S}$ is first passed through the ``filtering'' procedure: any stored cluster $s \in \mathcal{S}$ is removed if:
\begin{itemize}
    \item any of its constituents has disappeared, or
    \item any constituent has undergone a collision after the time when $s$ was registered (i.e., its recorded $t_{\text{coll}}$ is outdated compared to the current global state).
\end{itemize}

Two clusters $p$ and $s$ are overlapping if their constituent sets are not disjoint:
$$
p \sim s
\;\Leftrightarrow\;
C(p) \cap C(s) \neq \varnothing
$$
For each candidate $p \in \mathcal{P}$, the set of overlapping stored clusters is defined as:
$$
\mathcal{O}(p) = \{ s \in \mathcal{S} \mid p \sim s \}
$$

The update rule for $\mathcal{S}$ is as follows:
\begin{enumerate}
    \item If $\mathcal{O}(p) = \varnothing$, then $p$ is a new physical cluster:
    \begin{equation*}
      \mathcal{S} \gets \mathcal{S} \cup \{ p \}
    \end{equation*}
    \item If $\mathcal{O}(p) \neq \varnothing$, let
      \begin{equation*}
      s_{\max} = \arg\max_{s \in \mathcal{O}(p)} A(s)
      \end{equation*}
      be the largest previously registered variant of this identity.

      The candidate $p$ replaces all overlapping entries if and only if:
      \begin{equation*}
      A(p) > A(s_{\max})
      \end{equation*}
      In this case:
      \begin{equation*}
      \mathcal{S} \gets \bigl( \mathcal{S} \setminus \mathcal{O}(p) \bigr) \cup \{ p \}
      \end{equation*}
\end{enumerate}

This rule enforces two principles:
\begin{itemize}
    \item Preferential growth: a cluster that has accreted additional baryons is always accepted.
    \item Controlled dissociation: the cluster shrinkage is permitted only when supported by the evidence of re-scattering: increased collision time of at least one constituent.
\end{itemize}
The current rule is a conservative default. It can be easily adjusted or made stricter if needed for specific physics studies.

After applying the described procedure, baryons which artificially left previously ``stable'' clusters are returned back, however, there is a possibility, that at the current time step $t_{\text{step}}^{\text{curr}}$ these baryons of interest formed with other baryons new ``stable'' clusters. Since we consider the original ``stable'' cluster formed at some previous time step $t_{\text{step}}^{\text{prec}}$ decay as artificial, the question arises: could the new cluster be formed $t_{\text{step}}^{\text{curr}}$ without baryon which we remove back to the original cluster. To resolve this problem, the MST cluster finding procedure is applied again on top of the particles vector of the current time step after removing all baryons, bound into the stable cluster set $\mathcal{S}$. Then, if the newly found clusters satisfy the conditions of being ``stable'', they are added to $\mathcal{S}$. On the studied data set the application of the second MST procedure after each stable clusters tracking step lead to only $\sim 3\%$ increase of the overall stable cluster multiplicity. The most pronounced difference appears at the spectators region: forward and backward rapidity intervals close to the projectile and target beam rapidities, where fragments originate mainly from the non-overlap parts of the colliding nuclei rather than from the participant ``fireball''.

\begin{figure}[ht!]
    \includegraphics[width=\linewidth]{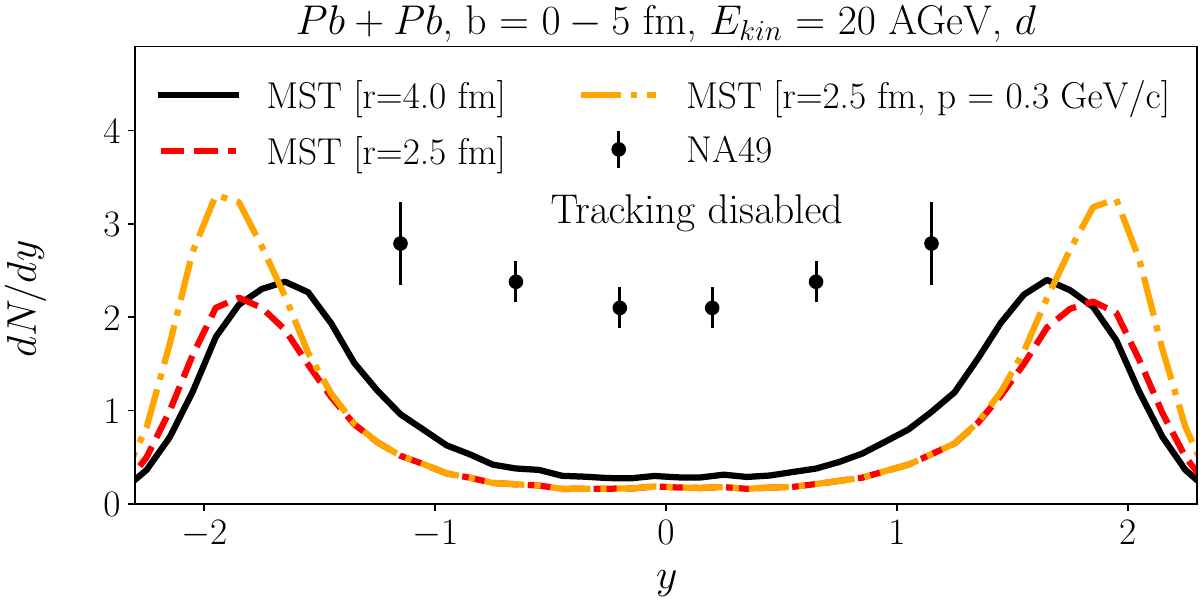}
    \includegraphics[width=\linewidth]{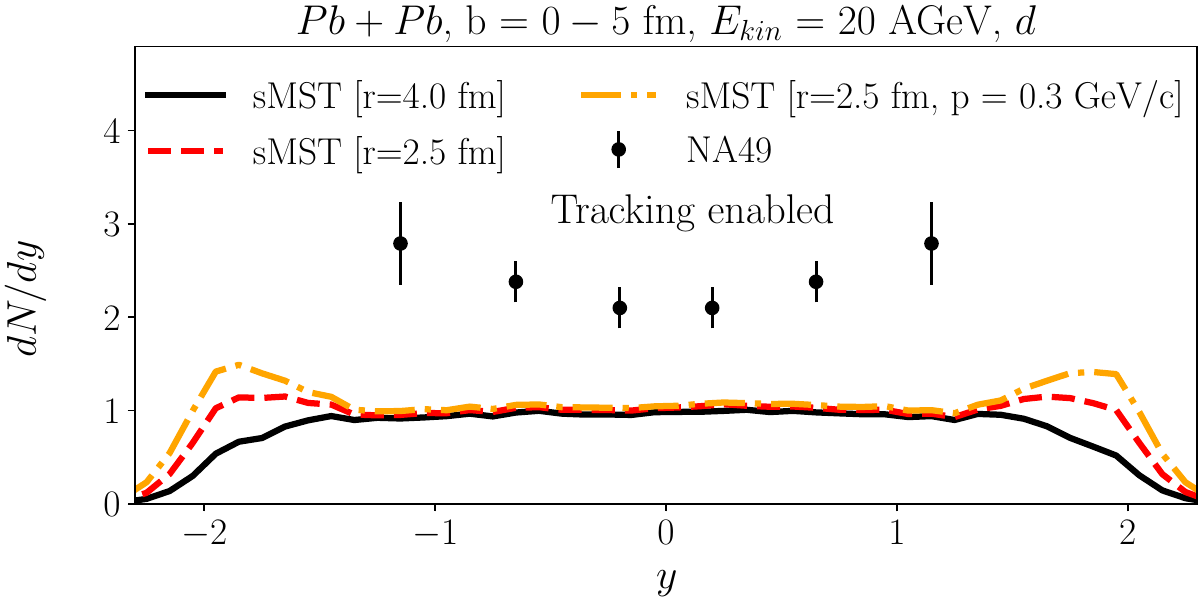}
    \includegraphics[width=\linewidth]{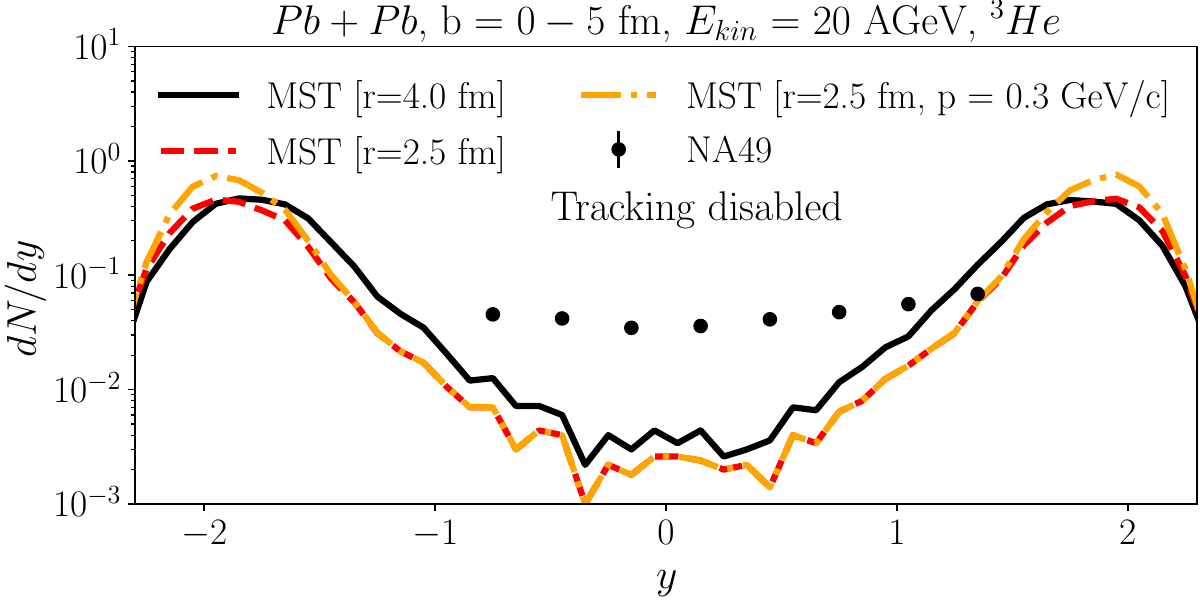}
    \includegraphics[width=\linewidth]{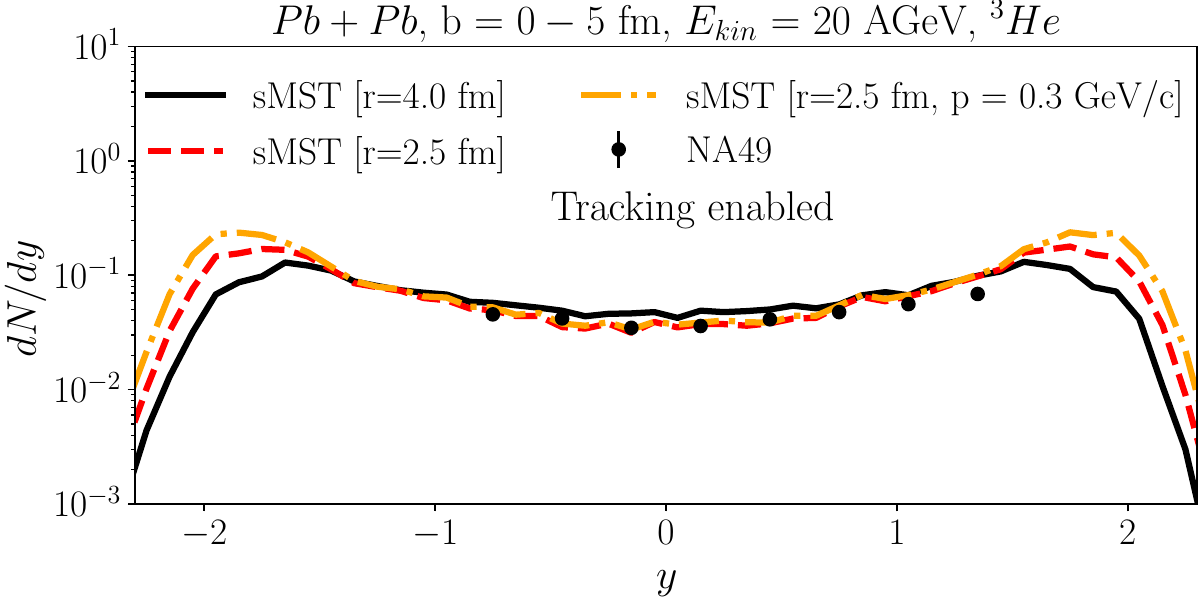}
    \caption{Rapidity density distribution of deuterons and $^3He$ with and without stable clusters tracking procedure with several parameters sets for the MST edges (\ref{tab:tracking}): black solid lines -- T1 set, T2 -- red dashed lines, T3 set -- orange dash-dotted lines. Black circles -- NA49 experimental data\cite{NA49:2016qvu}.}
\label{fig:tracking}
\end{figure}

After processing all $p \in \mathcal{P}$ within all subsequent time steps, the updated set $\mathcal{S}$ contains an ensemble of stable bound clusters at the end of reaction.
Fig. \ref{fig:tracking} illustrates the work of the stable cluster tracking algorithm which is denoted as ``sMST'' (``smart'' MST) on the plots. The parameters sets used to define the MST edges are shown in the Table \ref{tab:tracking}. On the two upper panels the rapidity density distributions for deuterons at the lowest NA49 experiment energy \cite{NA49:2016qvu} are shown, on the two bottom panels same distributions are presented for the $^3He$ correspondingly. The (s)MST procedure with T1 set is shown with black solid lines, (s)MST with T2 set is represented by the red dashed lines, T3 set -- by orange dash-dotted lines. In both cases when the tracking procedure was disabled, the clusters multiplicity in the central rapidity interval around $y = 0$ (mid-rapidity region) at the end of reaction are much lower than the experimental measurements. The enabling of the ``sMST'' tracking procedure drastically improves the mid-rapidity clusters stability.

\begin{table}[ht!]
\centering
\begin{tabular}{ccc}
\toprule
  & $\Delta r$, fm & $\Delta p$, GeV/c\\
\midrule
T1 & 4.0  & -- \\
T2 & 2.5  & -- \\
T3 & 2.5  & 0.3 \\
\bottomrule
\end{tabular}
\caption{MST parameters for the ``sMST'' tracking procedure.}
\label{tab:tracking}
\end{table}

\section{Coalescence}
\label{sec:Coal}
More than half of a century passed since the coalescence approach was firstly proposed \cite{Butler:1961pr} to describe the deuterons production in emulsions, then modified to describe the light nuclei production in relativistic heavy-ion reactions \cite{Gutbrod:1976zzr}. The CCL implements the 'box-like' coalescence proposed and successfully used by the authors of the UrQMD \cite{Bass:1998ca,Bleicher:1999xi} transport approach in \cite{Sombun:2018yqh, Reichert:2022mek}.

In this coalescence approach it is assumed that nuclei can be formed after the last collision of its constituent particles (protons, neutrons, hyperons) -- so called ``freeze-out'' condition. At the ``freeze-out'' times the relative distances between particles in the coordinate and momentum spaces $\Delta r$ and $\Delta p$ are calculated pairwise in the pair COM frame, the particle pair can potentially form nuclei if their $\Delta r < \Delta r_{\text{max}}$ and $\Delta p < \Delta p_{\text{max}}$. Then, next particle can be added to the pair by satisfying same conditions, assuming that the pair momentum and position are $\mathbf{p}_{\text{pair}} = \mathbf{p}_1 + \mathbf{p}_2$ and $\mathbf {r}_{\text{pair}} = (\mathbf{r}_1 + \mathbf{r}_2) / 2$. After collecting potential nuclei, the formation probability (spin-isospin factor) is applied.

\begin{figure}[ht!]
    \includegraphics[width=\linewidth]{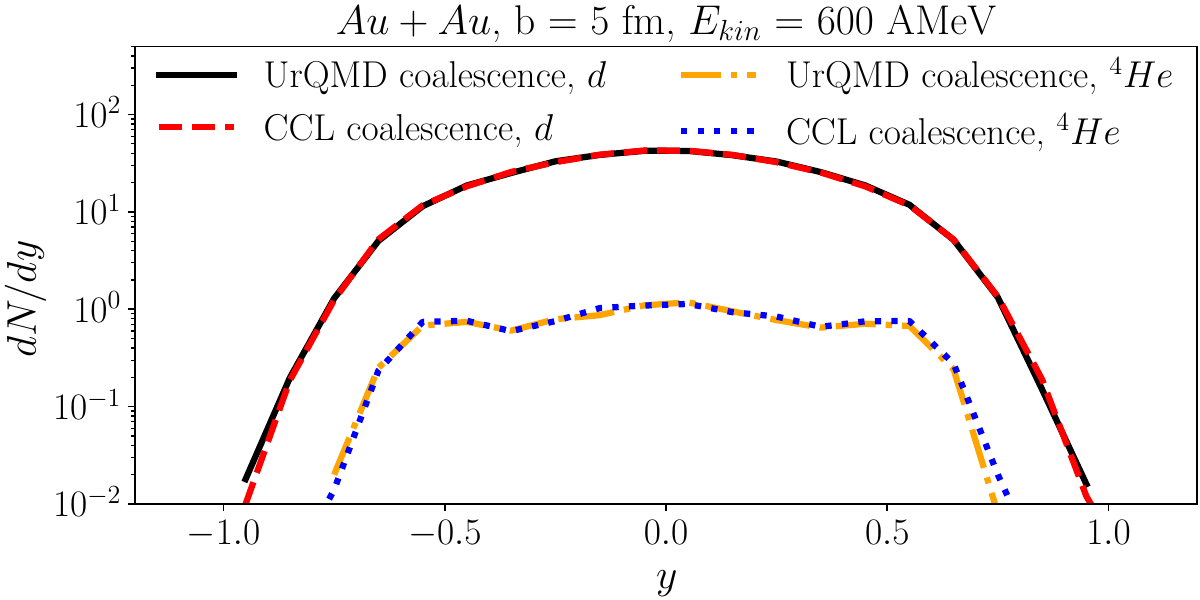}
    \caption{Rapidity density distribution.
    Black solid lines and orange dash-dotted lines -- $d$ and $^{4}He$ from UrQMD, red dashed lines and  blue dotted lines -- same from CCL coalescence.
    Probabilities to form nuclei are taken from the UrQMD model source code.
    }
    \label{fig:coal_alg}
\end{figure}

Currently, in the CCL the coalescence approach is applicable up to $^{4}He$ and $^{4}H_{\Lambda}$, the UrQMD built-in coalescence is advanced for anti-nuclei and $S = -2$ hypernuclei.
It is not possible to make absolutely equal port of the coalescence algorithm between two different approaches, however, the Fig. \ref{fig:coal_alg} shows almost perfect match between UrQMD built-in algorithm and the CCL coalescence implementation. Some discrepancy can also come from the fact that methods were applied to the different data sets, as the built-in coalescence takes out baryons bound in nuclei from the particles output.

\subsection{CCL coalescence procedure}
The coalescence of particles into light nuclei is determined by spatial and momentum criteria in the center-of-mass (COM) frame of particle pairs, with optional binding energy or probabilistic (e.g. spin-isospin factor) conditions for the final nuclei selection.

Like the MST-based clusterization, the coalescence procedure rely on the pairwise processing. The total energy and momentum of the particles pair are:
\begin{equation}
T_{\text{pair}} = E_1 + E_2
\end{equation}
Then, the pair velocity:
\begin{equation}
\beta_x = \frac{p_{x,1} + p_{x,2}}{T_{\text{pair}}}, \quad \beta_y = \frac{p_{y,1} + p_{y,2}}{T_{\text{pair}}}, \quad \beta_z = \frac{p_{z,1} + p_{z,2}}{T_{\text{pair}}},
\end{equation}
where $E_i$, $\mathbf{p}_i = (p_{x,i}, p_{y,i}, p_{z,i})$ are the energy and momentum of particle $i$ in the lab frame. Both particles are Lorentz-boosted to the pair's COM frame using the boost vector $\mathbf{\beta} = (\beta_x, \beta_y, \beta_z)$.

Then, to ensure that spatial coordinates are calculated at the same time, the particles positions are adjusted to a common latest ``freeze-out'' time within the pair. Let $t_1$ and $t_2$ be the ``freeze-out'' times of particles $p_1$ and $p_2$ in the pair COM frame, the time difference is $\Delta t = |t_1 - t_2|$. If $t_1 < t_2$, the particle $p_1$ position is propagated with straight lines to the time $t_2$:
\begin{gather*}
x_1 \gets x_1 + \frac{\Delta t}{E_1} p_{x,1}, \;
y_1 \gets y_1 + \frac{\Delta t}{E_1} p_{y,1}, \;
z_1 \gets z_1 + \frac{\Delta t}{E_1} p_{z,1},
\\
t_1 \gets t_2
\end{gather*}
If $t_2 < t_1$, adjust the $p_2$ particle position is propagates similarly.

At this moment particles coordinates in the pair COM frame are synchronized, and the proximity check can be performed. The squared distance between the particles in the synchronized COM frame is:
\begin{equation}
dr^2 = (x_1 - x_2)^2 + (y_1 - y_2)^2 + (z_1 - z_2)^2
\end{equation}
The particles pass the distance check if:
\begin{equation}
dr^2 \leq R^2,
\end{equation}
where $R$ is an algorithm control parameter -- coalescence radius.
When particles passes check for the proximity in the coordinate space, the squared relative momentum is computed (momenta are in the COM frame):
\begin{equation}
dp^2 = (p_{x,1} - p_{x,2})^2 + (p_{y,1} - p_{y,2})^2 + (p_{z,1} - p_{z,2})^2,
\end{equation}
The particles pass the momentum check if:
\begin{equation}
dp^2 \leq P^2,
\end{equation}
where $P$ is a maximum momentum difference -- an algorithm control parameter.
The particles can coalesce if both checks are satisfied. However, since only light nuclei formation is implemented in the coalescence procedure, the QMD approaches gives a multitude of that small clusters correlated in the coordinate-momentum space, moreover, some of these smaller clusters could be parts of bigger fragments. Thus, in CCL, after finding coalescence candidates, final nuclei are chosen by the probability driven by the spin-isospin coupling (see \cite{Sombun:2018yqh} for more details).
After passing the probability cuts, the final nuclei are collected.
For nuclei with more than two constituents, the procedure is applied iteratively: a pair which satisfies the coalescence criteria is first treated as a temporary composite object, then, the next candidate baryon is tested against this temporary object using the same spatial and momentum cuts.
This sequence is repeated until the required particle content of the candidate nucleus, for example $^3He$ or $^4He$, is reached.

Technically, the coalescence realized in a such way, that it can not be applied simultaneously for all implemented nuclei, but must be used in a ``chain'', finding heavier nuclei first and then moving to lighter:
\begin{equation*}
^4_\Lambda H \rightarrow ^4 He \rightarrow  ^3_\Lambda H \rightarrow t \rightarrow ^3 He \rightarrow d
\end{equation*}
Particles which enters to the final coalescence clusters are removed from the initial particles vector to ensure that they are not double-counted on the each subsequent coalescence step.

\subsection{Coalescence results}
The limited statistics of 50k central (b = $0 - 5$ fm) $Pb + Pb$ events simulated within PHQMD transport approach at $E_{\text{lab}} = 20$ AGeV was used to test the CCL coalescence procedure. Several sets of coalescence parameters ($\Delta r$, $\Delta p$) were used, all they are listed in the Table \ref{tab:cls_param}. The ``M1'' set parameters are imported from \cite{Reichert:2025rnw}, the ``M2'' parameters simulates the classical MST-based clusterization parameters, the ``M3'' are taken similar to those within the ``aMST'' procedure \cite{Coci:2023daq} (increased coalescence radius) and ``M4'' are the mix of all previous parameters chosen to better description of available data of the NA49 experiment at this energy (\cite{NA49:2016qvu}).

\begin{table*}[ht!]
\centering
\begin{tabular}{l * {10}{c}}
\toprule
 & \multicolumn{2}{c}{$d$} & \multicolumn{2}{c}{$t$/$^3He$} & \multicolumn{2}{c}{$^4He$} & \multicolumn{2}{c}{$^3_{\Lambda} H$} & \multicolumn{2}{c}{$^4_{\Lambda} H$} \\
\cmidrule(lr){2-3} \cmidrule(lr){4-5} \cmidrule(lr){6-7} \cmidrule(lr){8-9} \cmidrule(lr){10-11}
 & $\Delta r$ & $\Delta p$ & $\Delta r$ & $\Delta p$ & $\Delta r$ & $\Delta p$ & $\Delta r$ & $\Delta p$ & $\Delta r$ & $\Delta p$ \\
\midrule
M1 &
4.0 & 0.33 & 3.5 & 0.45 & 3.5 & 0.55 & 9.5 & 0.15 & 9.5 & 2.5 \\
M2 &
2.5 & 0.285 & 2.5 & 0.285 & 2.5 & 0.285 & 2.5 & 0.285 & 2.5 & 0.285  \\
M3 &
4.0 & 0.3 & 4.0 & 0.3 & 4.0 & 0.3 & 4.0 & 0.3 & 4.0 & 0.3 \\
M4 &
4.0 & 0.33 & 2.5 & 0.285 & 3.5 & 0.3 & 9.5 & 0.3 & 9.5 & 0.3 \\
\bottomrule
\end{tabular}
\caption{CCL parameter sets for coalescence and ``mixed'' procedure. $\Delta r$ are in fm, $\Delta p$ are in GeV/c.}
\label{tab:cls_param}
\end{table*}

\begin{figure}[ht!]
    \includegraphics[width=\linewidth]{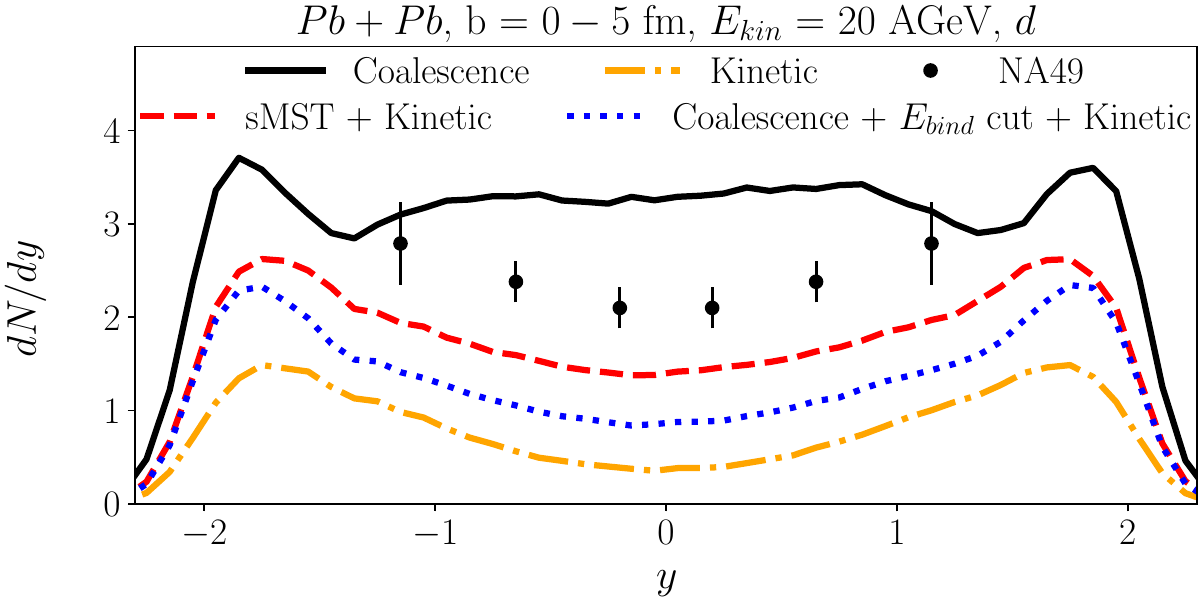}
    \caption{Rapidity density distribution of deuterons.
    Black solid lines are coalescence deuterons. The orange dash-dotted line are ``kinetic'' deuterons (see text). The red dashed lines shows the sum of the ``sMST'' and ``kinetic'' deuterons. Black circles -- NA49 experimental data (measured and mirrored) \cite{NA49:2016qvu}.}
    \label{fig:coal_mst_d}
\end{figure}

On the Fig. \ref{fig:coal_mst_d} rapidity distributions of deuterons are shown. Several mechanisms of deuteron formation (or finding) are presented on this figure:
\begin{itemize}
    \item Coalescence deuterons chosen with spin-isospin coupling criteria  -- solid black line. ``M1'' parameters set was used for $\Delta r$ and $\Delta p$. ``Coalescence'' on the plot.
    \item ``kinetic'' deuterons formed in $2 \leftrightarrow 3$ reactions through the pion catalyzer (\cite{Coci:2023daq}) -- orange dash-dotted line.
    \item The sum of ``kinetic'' and ``MST'' (or ``potential'') deuterons found by the MST-based cluster finding procedure with stable clusters tracking (``sMST'' with T2 parameters set) -- red dashed line.

\end{itemize}
As is seen on the plot, the coalescence approach based on the ``spin-isospin coupling'' not only overestimates the experimental data points, but also show the shape of the distribution with local maximum at mid-rapidity $y = 0$, which differs from the experimental data shape. The multiplicity of the ``kinetic'' deuterons produced in $2 \leftrightarrow 3$ reactions through the pion catalyzer (\cite{Coci:2023daq}) is rather small at this energy, however the final result must be the sum of ``kinetic'' deuterons and those found by (s)MST or formed via coalescence. The ``sMST + kinetic'' sum is describes the experimental data shape quite well, however, underestimates it. There is no reason to make that similar sum for the probabilistic coalescence as it already overestimate the data.

The coalescence results (solid black line) differs from those from \cite{Kireyeu:2023spj}, but this can be explained by the smaller coalescence parameters $\Delta r = 3.575$ fm and $\Delta p = 0.285$ GeV/c used in previous study, which led to the smaller yield of deuterons.

Additionally, coalescence deuterons after the negative binding energy selection combined with ``kinetic'' deuterons are plotted with blue dotted line. The negative binding energy $E_{\text{bind}} < 0$ selection is too strong and addition of the ``kinetic'' deuteron can not compensate the difference with the experimental data, however, fix the overall shape.

\begin{figure}[ht!]
    \includegraphics[width=\linewidth]{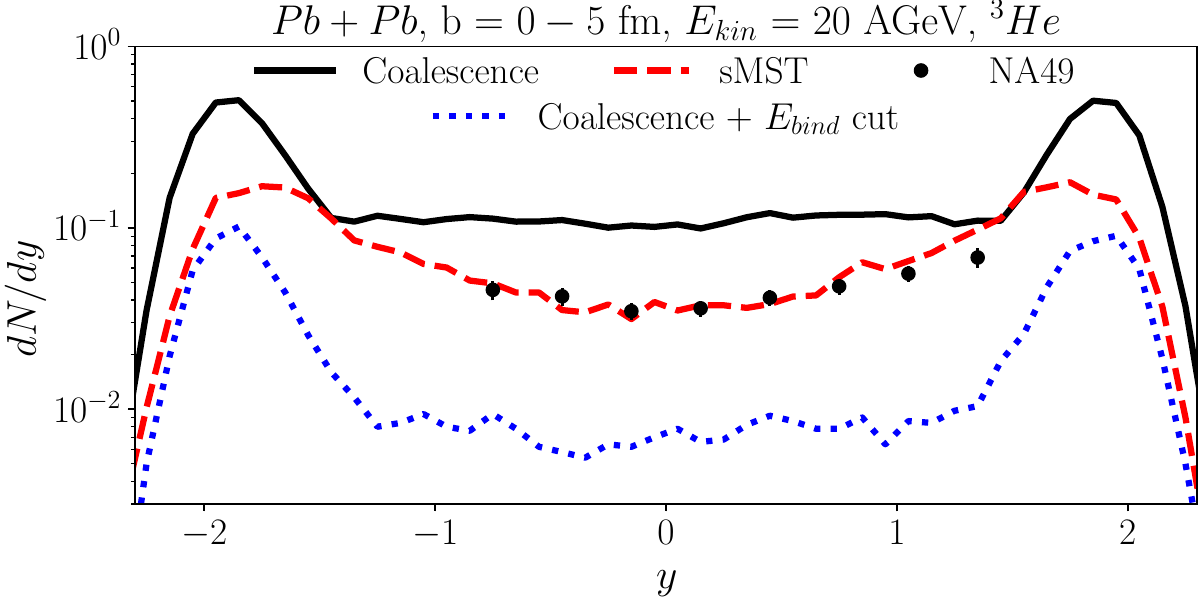}
    \caption{
    Rapidity density distribution of $^3He$.
    Black solid lines are coalescence with spin-isospin coupling cut. The red dashed lines shows the ``sMST'' results. Black circles -- NA49 experimental data\cite{NA49:2016qvu}.
    }
    \label{fig:coal_mst_he3}
\end{figure}

In Fig. \ref{fig:coal_mst_he3} similar results are presented for $^3He$ formed via coalescence (solid black line) or found by the MST procedure with additional ``sMST'' tracking with T2 parameters set (red dashed line). Here the situation is almost the same:
\begin{itemize}
    \item The coalescence results are not describing the experimental data shape and overestimates the data more than by the factor of 2 at the mid-rapidity region.
    \item The ``sMST'' results fits the data perfectly.
\end{itemize}
As on the Fig. \ref{fig:coal_mst_d}, coalescence with negative binding energy selection is also shown by the blue dotted line. Due to the limited statistics the distribution fluctuates in mid-rapidity $|y| < 1$, but overall conclusion is the same as for deuterons: the $E_{\text{bind}} < 0$ cut suppression for the coalescence nuclei is too strong.

\begin{figure}[ht]
    \includegraphics[width=\linewidth]{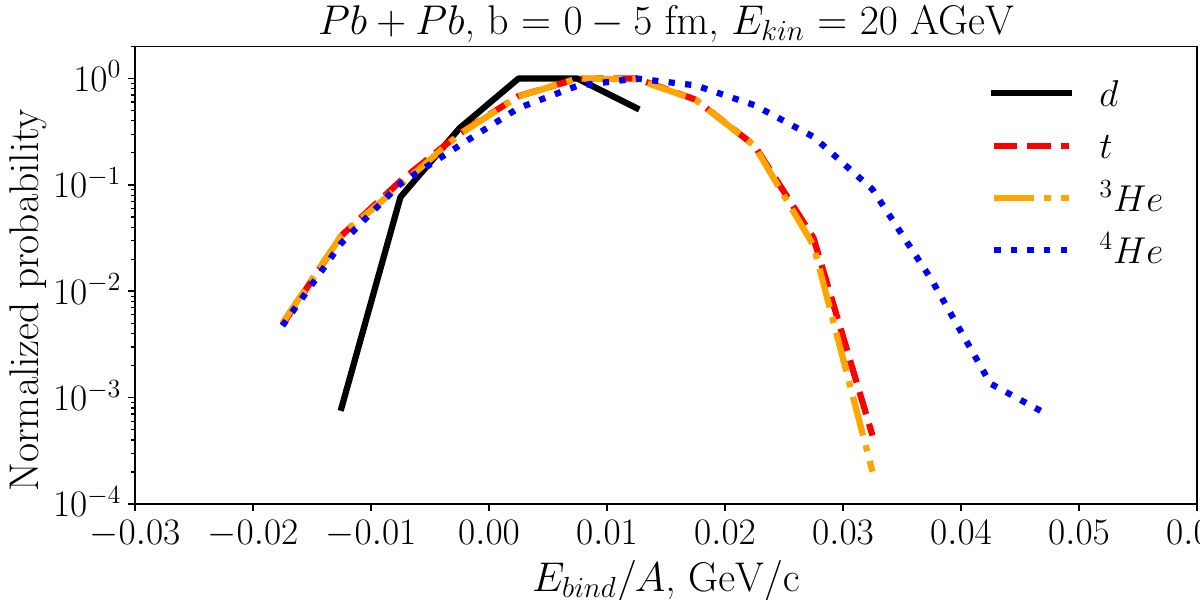}
    \caption{Binding energy per nucleon for found by the ``spin-isospin'' based coalescence nuclei: $d$ (black solid lines), $t$ (red dashed), $^3He$ (orange dash-dotted) and $^4He$ (blue dotted lines).}
    \label{fig:coal_prob_ebind}
\end{figure}

It is possible within the CCL library to check the binding energy of the produced ``Coalescence'' nuclei by the exactly same way used for SA procedure: the results for $d$, $t$, $^3He$ and $^4He$ are shown on the Fig. \ref{fig:coal_prob_ebind}. Curves on this plot are normalized to each to its maximum to be comparable. The picture is: more than 85\% of the each ``Coalescence'' nuclei would be considered as ``not bound'' within ``sMST'' or SA-based approaches as their binding energy $E_{\text{bind}} > 0$.

As there is possibility to check the binding energy of the coalescence nuclei, one can ask what will happen is we use the negative binding energy condition mixed with coalescence parameters $\Delta r$ and $\Delta p$. Even if in that case this ``mixed'' procedure should not be called ``Coalescence'' anymore, it is worth showing its results. One can expect that such ``mixed'' procedure must give results close to that from the MST-based clusterization after ``sMST'' tracking procedure, however, the ``freeze-out'' time synchronization can affect the particles distribution in the coordinate space.

\begin{figure}[ht!]
    \includegraphics[width=\linewidth]{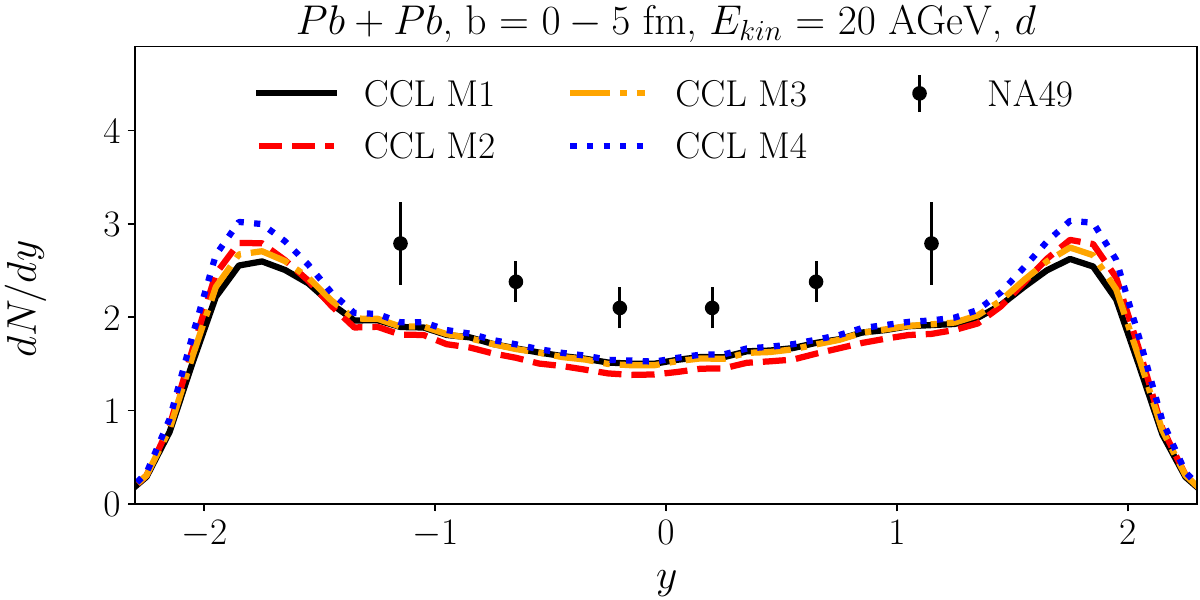}\\
    \includegraphics[width=\linewidth]{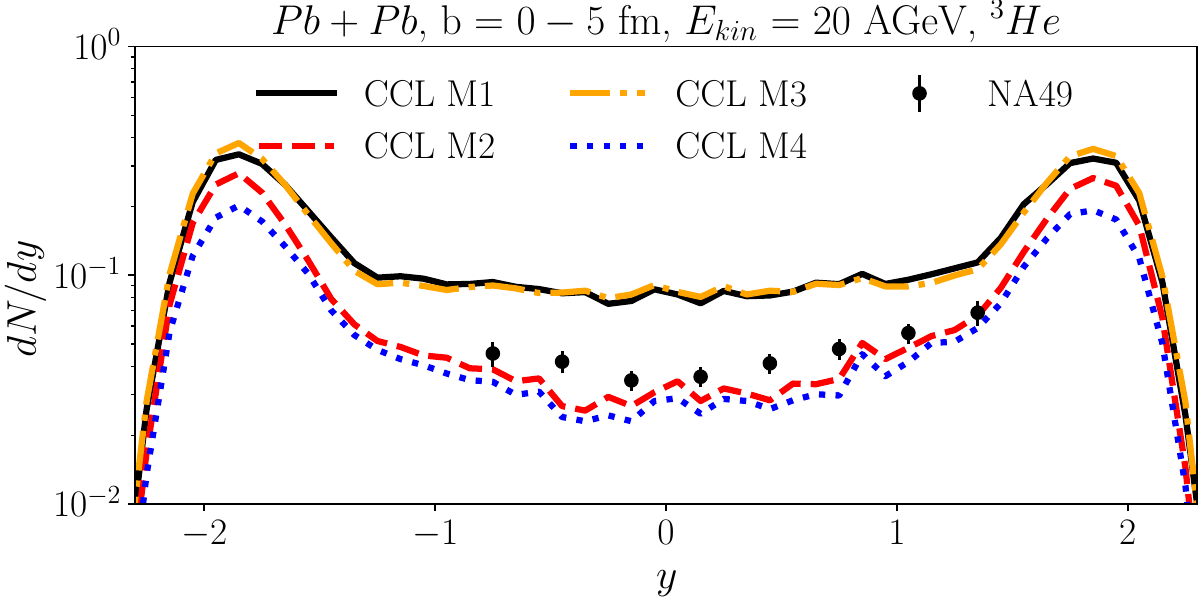}
    \caption{Rapidity density distribution of deuterons (top panel) and $^3He$ (bottom) found by the coalescence based on the $E_{\text{bind}} < 0$ cut. Deuteron lines are coupled with ``kinetic'' deuterons. Black solid lines are for the initial ``M1'' set of parameters, orange dash-dotted lines -- for the ``M2'' set,  ``M3'' set -- red dashed lines and blue dotted lines are for the set ``M4''.
    Black circles -- NA49 experimental data (measured and mirrored) \cite{NA49:2016qvu}.}
    \label{fig:coal_heavier}
\end{figure}

In Fig. \ref{fig:coal_heavier} the binding energy cut based ``mixed'' cluster finding procedure was tested with several parameters sets to find whose which allows for the better description of the experimental data. The spin-isospin combinatorial factor was excluded in this case. As on the Fig. \ref{fig:coal_mst_d}, all deuteron lines are coupled with ``kinetic'' deuterons. It is interesting, that with the binding energy selection criteria deuterons are not very sensitive to the different $\Delta r$ and $\Delta p$ parameters. This can be explained by the latest position of deuterons in the cluster finding chain: relatively loose in the coordinate space particles correlations are ``taken'' by heavier nuclei. The shape of the deuterons found by the ``mixed'' procedure differs from that obtained by the ``sMST'' procedure: more flat in the ``mixed'' case, and becomes closer to the data at larger rapidity in the case of the ``sMST''.

The helium-3 production is better described with the reduced coalescence radius $\Delta r = 2.5$ fm. Another interesting observation concerning $^3He$ is the difference in the shape within $|y| \lesssim 1.3$ -- it became similar to the experimental rapidity distribution with lower $\Delta r$ and 'plateau'-like with bigger $\Delta r = 3.5$ or $\Delta r = 4.0$ fm when $^3He$ found by ``sMST'' on the Fig.\ref{fig:coal_mst_he3} has similar shape as the experimental data using the same $\Delta r$. It is also visible, that the ``mixed'' case with smaller $\Delta p < 0.285$ GeV is even closer to the data than ``sMST'' without momentum cuts. This observation can suggest to include particles momentum checks for further studies with (s)MST cluster finding procedure.

These are the first results of the procedures included in the CCL library, thus, not all nuclei at all energies are tested, however, is it clearly seen that this library has some potential for the future use fast tests and parameters tuning in the already well established procedures.

\section{Summary}
Seven cluster finding procedures were tested in this work: full-graph MST, DSU-based MST, SACA, ``CCL SA'' chain, coalescence based on the ``spin-isospin'' combinatorial factor, ``box-like'' coalescence joined with the $E_{\text{bind}} < 0$ cut and finally ``mixed'' procedure based on the coalescence algorithm and the negative binding energy selection criteria''.

\begin{itemize}
    \item The MST-based clusterization reproduce ALADIN observables quite well, however, the better description, which is achieved at the very late time step $t = 200$ fm/c is still overestimate the data points on the ``top'' of the ``rise and fall'' curve.

    \item The application of the SACA and/or SA procedures reduces the total system $E_{\text{bind}}$ at earlier times (finds more ``bound'' systems) and yields into the better agreement with the experimental data.

    \item At the later times ($t = 100$ fm/c and higher) the mean binding energy of the system $\langle E_{\text{bind}} \rangle$ for the MST-based clusterization is similar to the SACA and SA procedures. This could be explained by the fact that at the later times the system is so dilute, so baryons with high relative velocities/momentum are well separated in space, so the found by the proximity criteria (MST) clusters can be already bound enough. That fact also leads to the priority of the DSU-based MST clusterization at later times over the SACA/SA due to negligible computational time required by the DSU-MST.

    \item Due to the differences between SACA and CCL SA procedures, their results are not identical at any time step, however, still very close to each other.

    \item At low energies as $E_{lab} = 600$ MeV the control parameters of the cluster finding and $E_{\text{bind}}$ minimization procedures must not be fixed, but chosen accordingly to the system evolution.

    \item The stable clusters tracking procedure ``smart MST'' (``sMST'') allows for the noticeable reparation of the artificially ``lost'' nuclei (due to QMD operation features). Proposed procedure works on-line, does not requires the knowledge of the last collision time of the particles and thus can be embedded to the transport codes. The sensitivity of the cluster tracking procedure to the ``MST'' edges definition was evaluated.

    \item The coalescence based on the spin-isospin combinatorial factor can describe the experimental data, however, more than than 85\% of found nuclei are not bound with positive binding energy $E_{\text{bind}} > 0$. Addition of the negative binding energy criterion to the ``box-like'' coalescence strongly suppress the yields of nuclei.

    \item The ``mixed'' procedure, which combines the coalescence algorithm steps and the negative binding energy condition instead of the spin-isospin combinatorial factor does not fully reproduce the MST (``sMST'') results as one could expect, however, gives some insights on the possible parameters tuning for the MST-based cluster finding procedures at relatively high energies of $E_{\text{kin}} = 20$ AGeV.

    \item The Helium-3 nuclei found by the MST-based cluster finding procedure (``sMST'') and the coalescence algorithm applied on top of the exactly the same set of baryons was presented for the first time and compared to the NA49 experimental data. Results favors the ``sMST'' procedure.
\end{itemize}

\section{Acknowledgments}
We would like to acknowledge the theoretical support and extensive discussions with J\"org Aichelin and Elena Bratkovskaya, without which this work could not have been completed.
We would also like to thank Marcus Bleicher and Jan Steinheimer for the insights on the UrQMD internal coalescence algorithm and its parameters.

Also we acknowledge the discussions with Arnaud Le F\`evre on the FRIGA approach.

Furthermore, we acknowledge help for the independent cross-checks of the obtained results by Jiaxing Zhao and Vadim Voronyuk.

\bibliography{references}

\end{document}